\newcommand{\std}[1]{
  \usepackage{amssymb}
  \usepackage{amsmath}
  \usepackage[final]{graphicx}
  \usepackage{a4,a4wide}
  \renewcommand{\baselinestretch}{#1}
  \parindent 0pt
  \parskip 2ex
  \pagestyle{headings}
  \begin{document}
}
\newcommand{\article}[1]{
  \documentclass[12pt,fleqn]{article}\std{#1}
}
\newcommand{\artikel}[1]{
  \documentclass[12pt,twoside,fleqn]{article}\usepackage{german}\std{#1}
}
\newcommand{\revtex}{
  \documentstyle[preprint,aps,eqsecnum,amssymb,fleqn]{revtex}
  \begin{document}
}
\newcommand{\book}[1]{
  \documentclass[12pt,twoside,fleqn]{book}
  \addtolength{\oddsidemargin}{2mm}
  \addtolength{\evensidemargin}{-4mm}
  \std{#1}
}
\newcommand{\buch}[1]{
  \documentclass[10pt,twoside,fleqn]{book}
  \usepackage{german}
  \usepackage{makeidx}
  \makeindex
  \std{#1}
}
\newcommand{\foils}[1]{
  \documentclass[fleqn]{article}
  \usepackage{amssymb}
  \usepackage{amsmath}
  \usepackage[final]{graphicx}
  \renewcommand{\baselinestretch}{#1}
  \setlength{\voffset}{-3cm}
  \setlength{\hoffset}{-4cm}
  \setlength{\textheight}{27cm}
  \setlength{\textwidth}{19cm}
  \parindent 0ex
  \parskip 0ex 
  \pagestyle{plain}
  \begin{document}
  \huge
}
\newcommand{\landfoils}[1]{
  \documentclass[fleqn]{article}
  \usepackage{amssymb}
  \usepackage{amsmath}
  \usepackage[final]{graphicx}
  \renewcommand{\baselinestretch}{#1}
  \setlength{\hoffset}{-5cm}
  \setlength{\voffset}{-1.5cm}
  \setlength{\textwidth}{27cm}
  \setlength{\textheight}{19cm}
  \parindent 0ex
  \parskip 0ex 
  \pagestyle{plain}
  \begin{document}
  \huge
}

\newcommand{\headline}[1]{\footnotetext{\sc Marc Toussaint, \today
    \hspace{\fill} file: #1}}

\newcommand{\sepline}{
\begin{center} \begin{picture}(200,0)
  \line(1,0){200}
\end{picture}\end{center}
}

\newcommand{\intro}[1]{\textbf{#1}\index{#1}}

\newtheorem{defi}{Definition}
\newenvironment{definition}{
\begin{quote}
\begin{defi}
}{
\end{defi}
\end{quote}
}

\newenvironment{block}{
\begin{quote} \begin{picture}(0,0)
        \put(-5,0){\line(1,0){20}}
        \put(-5,0){\line(0,-1){20}}
\end{picture}

}{

\begin{picture}(0,0)
        \put(-5,5){\line(1,0){20}}
        \put(-5,5){\line(0,1){20}}
\end{picture} \end{quote}
}

\newenvironment{summary}{
\begin{center}\begin{tabular}{|l|}
\hline
}{\\
\hline
\end{tabular}\end{center}
}

\newcommand{\inputReduce}[1]{
  
  {\sc\hspace{\fill} REDUCE file: #1}
}
\newcommand{\inputReduceInput}[1]{
  
  {\sc\hspace{\fill} REDUCE input - file: #1}
  \input{#1.tex}
}
\newcommand{\inputReduceOutput}[1]{
  
  {\sc\hspace{\fill} REDUCE output - file: #1}
}

\newcommand{\macros}{
  \newcommand{\0}{{\hat 0}}
  \newcommand{\1}{{\hat 1}}
  \newcommand{\2}{{\hat 2}}
  \newcommand{\3}{{\hat 3}}
  \newcommand{\5}{{\hat 5}}
  \newcommand{\QQ}{{\cal Q}}

  \renewcommand{\a}{\alpha}
  \renewcommand{\b}{\beta}
  \renewcommand{\c}{\gamma}
  \renewcommand{\d}{\delta}
    \newcommand{\D}{\Delta}
    \newcommand{\e}{\epsilon}
    \newcommand{\g}{\gamma}
    \newcommand{\G}{\Gamma}
  \renewcommand{\l}{\lambda}
  \renewcommand{\L}{\Lambda}
    \newcommand{\m}{\mu}
    \newcommand{\n}{\nu}
    \newcommand{\N}{\nabla}
  \renewcommand{\k}{\kappa}
  \renewcommand{\o}{\omega}
  \renewcommand{\O}{\Omega}
    \newcommand{\p}{\varphi}
  \renewcommand{\r}{\varrho}
    \newcommand{\s}{\sigma}
    \newcommand{\Si}{\Sigma}
  \renewcommand{\t}{\theta}
    \newcommand{\T}{\Theta}
  \renewcommand{\v}{\vartheta}
    \newcommand{\Y}{\Upsilon}

  \newcommand{\C}{{\bf C}}
  \newcommand{\R}{{\bf R}}
  \newcommand{\Z}{{\bf Z}}

  \renewcommand{\AA}{{\cal A}}
  \newcommand{\GG}{{\cal G}}
  \newcommand{\TT}{{\cal T}}
  \newcommand{\EE}{{\cal E}}
  \newcommand{\HH}{{\cal H}}
  \newcommand{\II}{{\cal I}}
  \newcommand{\KK}{{\cal K}}
  \newcommand{\MM}{{\cal M}}
  \newcommand{\CC}{{\cal C}}
  \newcommand{\PP}{{\cal P}}
  \newcommand{\RR}{{\cal R}}
  \newcommand{\YY}{{\cal Y}}
  \newcommand{\SOSO}{{\cal SO}}
  \newcommand{\GLGL}{{\cal GL}}

  \newcommand{\NNN}{\mathbb{N}}
  \newcommand{\ZZZ}{\mathbb{Z}}
  \newcommand{\RRR}{\mathrm{I\!R}}
  \newcommand{\CCC}{\mathbb{C}}
  \newcommand{\one}{{\bf 1}}

  \newcommand{\<}{\langle}
  \renewcommand{\>}{\rangle}
  \newcommand{\tr}{{\rm tr}}
  \newcommand{\lag}{\mathcal{L}}
  \newcommand{\inn}{\rfloor}
  \newcommand{\lie}{\pounds}
  \newcommand{\speer}{\parbox{0.4ex}{\raisebox{0.8ex}{$\nearrow$}}}
  \renewcommand{\dag}{ {}^\dagger }
  \newcommand{\h}{{}^\star}
  \newcommand{\w}{\wedge}
  \newcommand{\ow}{\stackrel{\circ}\wedge}
  \newcommand{\feed}{\nonumber \\}
  \newcommand{\comma}{\; , \quad}
  \newcommand{\period}{\; . \quad}
  \newcommand{\del}{\partial}
  \newcommand{\point}{$\bullet~~$}
  \newcommand{\doubletilde}{
  ~ \raisebox{0.3ex}{$\widetilde {}$} \raisebox{0.6ex}{$\widetilde {}$} \!\!
  }
  \newcommand{\topcirc}{\parbox{0ex}{~\raisebox{2.5ex}{${}^\circ$}}}
  \newcommand{\sym}{\topcirc}

  \newcommand{\half}{\frac{1}{2}}
  \newcommand{\third}{\frac{1}{3}}
  \newcommand{\fourth}{\frac{1}{4}}

}

\newcommand{\tmp}{\fbox{?}}
\newcommand{\Label}[1]{\label{#1}\fbox{\tiny #1}}

\macros
\newcommand{\path}{./}
\newcommand{\basepath}{./}
\newcommand{\setpath}[1]{
  \renewcommand{\path}{#1}
  \renewcommand{\basepath}{#1}}
\newcommand{\pathinput}[2]{
  \renewcommand{\path}{\basepath #1}
  \input{\path #2} \renewcommand{\path}{\basepath}}

\newcounter{parac}
\newcommand{\para}{\refstepcounter{parac}[\emph{\arabic{parac}}]~~}
\newcommand{\Pref}[1]{[\emph{\ref{#1}}\,]}
\article{1}
\renewcommand{\Label}[1]{\label{#1}}

\title{A gauge theoretical view of the charge concept in Einstein gravity
\footnote{Presented at the annual meeting of the German Physical Society -- Heidelberg, March 1999.}}
\author{
Marc Toussaint\\
\small Institute for Theoretical Physics, University of Cologne\\
\small 50923 K\"oln, Germany\\
\small {\tt www.thp.uni-koeln.de/\~{}mt/}}
\date{}

\maketitle
\begin{abstract}
  We will discuss some analogies between internal gauge theories and
  gravity in order to better understand the charge concept in gravity.
  A dimensional analysis of gauge theories in general and a strict
  definition of \emph{elementary}, \emph{monopole}, and
  \emph{topological} charges are applied to electromagnetism and to
  teleparallelism, a gauge theoretical formulation of Einstein
  gravity.

  As a result we inevitably find that the gravitational coupling
  constant has dimension $\hbar/\ell^2$, the mass parameter of a
  particle dimension $\hbar/\ell$, and the Schwarzschild mass
  parameter dimension $\ell$ (where $\ell$ means \emph{length}). These
  dimensions confirm the meaning of mass as elementary and as monopole
  charge of the translation group, respectively.  In detail, we find
  that the Schwarzschild mass parameter is a \emph{quasi-electric
    monopole charge of the time translation} whereas the NUT parameter
  is a \emph{quasi-magnetic monopole charge of the time translation}
  as well as a topological charge.  The Kerr parameter and the
  electric and magnetic charges are interpreted similarly. We conclude
  that each elementary charge of a Casimir operator in the gauge group
  is the source of a (quasi-electric) monopole charge of the
  respective Killing vector.

  Keywords: gauge theory of gravity, Kaluza-Klein, charge, monopole,
  mass, Taub-NUT.

\end{abstract}

\newpage
\parskip 0ex
\tableofcontents
\parskip 2ex

\section{Introduction}

In the fifties, Yang and Mills \cite{yangM} for the first time
formulated the $SU(2)$-gauge theory by strictly keeping to the
electromagnetic paradigm. At about the same time, Utiyama
\cite{utiyama} formulated the general gauge theory of a semi-simple
Lie group. These theories, as they explain the electro-weak and strong
\emph{forces}, were supplemented by the great success of
\emph{particle} physics to classify all leptons as representations of
the electro-weak symmetry and all hadrons as representations of the
flavor symmetry.  O'Raifeartaigh \cite{oraif97} gives a more detailed
insight into the history of gauge theories. The great success of such
theories has also influenced modern formulations of gravity -- one of
the four fundamental forces which should also be representable in the
framework of gauge theory.  However, the obvious difference between
the \emph{external} spacetime symmetries and \emph{internal}
symmetries (as considered by Yang and Mills) causes some difficulties
for a uniform formulation of all forces. Some ad-hoc assumption (the
\emph{soldering}) solves basic problems but perhaps diminishes the
beauty of the theory. We refer to \cite{hehlMMN} (more introductory
\cite{gron}) as a general formulation of gravity as a gauge theory
(see table \ref{Tgauge}). For this work it is most important to
understand teleparallelism as a gauge theory of translations with the
anholonomic coframe $\v^\a$ as gauge potential and torsion $T^\a$ as
field strength. With a specific lagrangian, this theory is equivalent
to Einstein gravity. This will enable us to reformulate standard
Einsteinian solutions in the framework of teleparallelism and thus to
interpret the solution parameters as translational charges.

Now, what is a charge? In general is seems plausible to define a
charge to be a specific and invariant property of a particle (usually
given by one number, perhaps an integer). Since in gauge theories we
take particles to be elements in a representation of the symmetry, we
are directly led to the most basic notion of a charge, the
\emph{elementary charge}, classifying the representation of the
symmetry the particle is an element of. But also a specific property
of the gauge field which a particle \emph{necessarily} induces may be
considered as a charge. Such is, e.g., the monopole character of the
electromagnetic field around an electron. This field is induced by the
coupling of the electron's elementary charge to the gauge field. Such
could also be the magnetic monopole character of, say, the
electromagnetic field around a Dirac monopole. But, since there exists
no magnetic-type elementary charge, there is no reason a for particle
induce such a field -- except for topology. We will see that in the
bundle formalism topological effects also motivate this third,
topological kind of charge, including the quasi-magnetic monopole
charge.

In the following we define these three kinds of charges and apply the
definitions on Taub-NUT and Kerr-Newman type solutions of
teleparallelism. It will be very satisfying to recognize the
Schwarzschild mass parameter as a \emph{quasi-electric monopole charge
  of the time translation} and the NUT parameter as a
\emph{quasi-magnetic monopole charge of the time translation}. The
Kerr parameter is interpreted similarly. These results are in perfect
analogy to monopoles in electromagnetism, they shed light on the
dimensions of parameters, and they emphasize the analogy between
internal and external gauge theories.

Before, in section 2, we insert a brief dimensional analysis of gauge
theories in general. The Kaluza-Klein formulation of electromagnetism
makes a comparison with gravity very simple.

\begin{table}[t]\center
\begin{tabular}{|c|c|c|c|}
  \hline
\raisebox{-2ex}{}theory & gauge group & connection & field strength \\ \hline
general gauge theory & semi-simple Lie group $G$ & $A \in \L^1(M,\GG)$ & $F=D\G \in \L^2(M,\GG)$ \\
electrodynamics & $U(1)$ & $A$ & $F=dA$ \\
(\emph{non-physical}) & affine group & $\doubletilde \G$ & $\doubletilde \! R$ \\
affine gauge theory & soldered affine group & $\G_\a{}^\b$, $\v^\a$ & $R_\a{}^\b$, $T^\a$ \\
teleparallelism & soldered translations & $\v^\a$ & $T^\a = d\v^a$ \\
\hline
\end{tabular}
\caption{
  Gravity may be described by formulating a gauge theory of the affine
  group. However, one has to ensure that the group, i.e.\ the
  Lie-algebra valued connection, applies to spacetime -- is
  \emph{soldered} to spacetime. This is done by splitting the
  connection into a linear part $\G_\a{}^\b$ (with matrix indices
  ${}_\a{}^\b$ that work on the basis $e_\a$ of the local tangent
  space) and an inhomogeneous part $\v^\a$ (that replaces the
  holonomic coframe $dx^\a$ and thereby realizes a translational
  gauge). The field strength splits into the curvature $R_\a{}^\b$ and
  the torsion $T^\a$. Discarding the linear gauge ($\G_\a{}^\b \equiv
  0$), the theory reduces to teleparallelism.}
\Label{Tgauge}
\end{table}

\section{Dimensional analysis of gauge theories}

The essential fields involved in a gauge theory of a Lie group $G$
(with algebra $\GG$) are the connection $A$, the field strength $F$,
the excitation $H$, the lagrangian $\lag$, and the Noether current
$\Si$.  From a geometrical point of view, the connection is introduced
as a $\GG$-valued 1-form on the principle bundle or, locally, as a
$\GG$-valued 1-form on spacetime, i.e.\ $A \in \L^1(M,\GG)$. It yields
the covariant exterior derivative $D=d+A$.

By its very definition, the exterior differentiation operator $d$ is
dimensionless, $[d] = 1$. Hence we also require the connection to be
dimensionless, $[A] = 1$. Now we need to give \emph{exactly two}
definitions in order to find all the remaining dimensions.  First, we
\emph{choose to define} the dimension of a lagrangian $\lag$ to be
$[\lag] = \hbar$. In the context of a classical gauge theory $\hbar$
is merely a name of a dimension as introduced here. However, thinking
of Huygen's principle and the path integral method, one may also call
$\hbar$ a \emph{phase$/2\pi$ unit}. And second, we define the basis
elements $\l_a$ of the algebra $\GG$ to have the dimension
$[\l_a]=g/\hbar$. Again, so far $g$ is merely a name of a dimension
introduced here. However, in the case of electromagnetism, it may be
replaced by the \emph{unit} $e$. Now it is easy to display the
dimensions of the components of $A \equiv A^a\, \l_a \equiv A_i{}^a\,
\l_a\, dx^i$ and $F \equiv F^a\, \l_a \equiv \half F_{ij}{}^a\, \l_a\, dx^i
\w dx^j$. You will find them in table \ref{Tdim}.

In Yang-Mills theories a lagrangian typically describes propagating
gauge fields, i.e.\ it is proportional to a square term of $F$. Here,
for generality, we only assume $\lag = \langle F \w H \rangle = F^a \w
H_a$, where we introduced the excitation $H$, which is a $\GG$-valued
2-form, and the metric $\langle \; , \; \rangle$ in $\GG$. We read off
the dimension of the excitation $H \equiv H^a\, \l_a \equiv \half
H_{ij}{}^a\, \l_a\, dx^i \w dx^j$ and of the Noether current $\Si_a :=
\d \lag / \d A^a$. For consistency, the dimension of the metric has to
be $[\langle \; , \; \rangle] = \hbar^2/g^2$. It follows $[\langle
\l_a,\l_b \rangle]=1$.  The dimension of $[H]/[F] = g^2/\hbar$ may be
interpreted as the dimension of the coupling constant $1/\k$ of a
dynamical lagrangian with $H \approx 1/\k\ \h F$.

In the case of electrodynamics, we only have one index $a = 0$ and we
set $[\l_0] = e/\hbar$. We see that the \emph{algebra components}
$F^a$, $H^a$, and $\Si_a$ carry the conventional dimensions, whereas
the dimensions of the fields $F$, $H$, $\Si$ are more unfamiliar. In
the case of a translational gauge theory, we assign the dimension
$1/\ell$ to the generators (where $\ell$ means \emph{length}) and find
that $[1/\k] = \hbar / \ell^2$. Since this dimensionality includes a
length dimension perturbation theory does not work. When embedding
electrodynamics in an extra dimension \`a la Kaluza-Klein, the $U(1)$
gauge is directly represented by the translation along the 5th
dimension. We can introduce a length \emph{unit} $\ell_5$ of this 5th
dimension by identifying $e/\hbar = 1/\ell_5$. This is a geometrical
interpretation of the electric unit $e$ as \emph{phase$/2\pi$ per
  length}. Besides, if the 5th dimension is $U(1)$ with perimeter
$L_5$, it seems natural that this `phase$/2\pi$ per length'-unit $e$
is quantized in quanta of $\hbar / L_5$. Thus, we may assume that the
perimeter of $U(1)$ is $L_5=\ell_5=\hbar/e$.

\begin{table}[t]\center
\begin{tabular}{|l|c|c|c|}
\hline
\raisebox{-3ex}{}& in general & in electrodynamics & \raisebox{1ex}{\parbox{18ex}{\center in translational gauge theories}} \\
\hline
$[\lag:=F^a \w H_a]$ \hspace{\fill} =: & $\hbar$ & Wb C & $\hbar$ \\
$[\l_a]$ \hspace{\fill} =: & $g/\hbar$ & 1/Wb & $1/\ell$ \\ 
\hline
$[A=A^a \l_a]$ \hspace{\fill} $\equiv$\; & 1 & 1 & 1\\
$[F=F^a \l_a]$ & 1 & 1 & 1\\
$[H=H^a \l_a]$ & $g^2/\hbar$ & C/Wb & $\hbar/\ell^2$ \\ 
\hline
$[A^a=A_i{}^a\, dx^i]$ & $\hbar/g$  & Wb & $\ell$ \\
$[F^a=\half\, F_{ij}{}^a\, dx^i \w dx^j]$ & $\hbar/g$ & Wb & $\ell$ \\
$[H^a=\half\, H_{ij}{}^a\, dx^i \w dx^j]$ & $g$ & C & $\hbar/\ell$ \\
$[\Si_a=\d\lag / \d A^a]$ & $g$ & C & $\hbar/\ell$ \\
\hline
$[\<\; , \;\>]=1/[\l]^2$ & $\hbar^2/g^2$ & Wb$^2$ & $\ell^2$ \\
$[\< F , H\>]$ & $\hbar$ & Wb C & $\hbar$ \\
$[1/\k]=[H]/[F]=[H^a]/[F^a]$ & $g^2/\hbar$ & C/Wb & $\hbar/\ell^2$ \\ 
\hline
$[\EE]=[\MM]=[F]$ & 1 & 1 & 1 \\
$[\EE^a]=[\MM^a]=[F^a]$ & $\hbar/g$ & Wb & $\ell$ \\
$[\II]$ & $g$ & C & $\hbar/\ell$ \\
\hline
\end{tabular}
\caption{
  The table displays the dimensions of essential fields and objects
  involved in a gauge theory. In particular, it gives the SI-units in
  the case of electrodynamics and the dimensions for a translational
  gauge theory. We stress that the first two rows in this table are
  definitions, the third is an identity, and the rest is a
  consequence. The last block includes the dimensions of monopole and
  topological charges.  The SI-units used in electrodynamics are
  C=Coulomb and Wb=Weber. We have the lagrangian $\lag$, group
  generators $\l_a$, gauge potential $A$, field strength $F$,
  excitation $H$, Noether current $\Si$, algebra metric $\<\; , \;\>$,
  coupling constant $1/\k$, quasi-electric and -magnetic charge $\EE$
  and $\MM$, and elementary charge $\II$.}
\Label{Tdim}
\end{table}

We want to point out again that any dimensional system of a gauge
theory (as long as all generators have the same dimension) may be
spanned by exactly two definitions, e.g.\ those for $[\lag]$ and
$[\l_a]$. This is the reason why every column in table \ref{Tdim}
includes exactly two dimensions (or units).

Finally we note that the dimension of the hodge star $\h$ in $n$
dimensions when applied on a $p$-form is $[\h]=l^{n-2p}$.

\section{Charge definitions}\Label{Cintcharges}

\subsection{Monopole charges}

We start by defining two types of monopole charges. These are
properties of the gauge configuration given by the gauge field
strength $F$:
\begin{align}
&\EE := \lim_{r\to\infty}\frac{1}{4\pi}\_{S^{n-2}(r)}\oint \h F
   && \text{\em quasi-electric monopole charge,} \Label{elcharge}\\
&\MM := \lim_{r\to\infty}\frac{1}{4\pi}\_{S^2(r)}    \oint F
   && \text{\em quasi-magnetic monopole charge.} \Label{magcharge}
\end{align}
The motivation for the definition of $\EE$ is obvious from the analogy
to Maxwell's inhomogeneous equation. The definition of $\MM$ may be
motivated by including magnetic charges in Maxwell's theory. Usually
this is done by modifying the homogeneous Maxwell equation and
introducing a source term on its rhs: $dF=\rho_{\rm mag}$. But we
think it is preferable to interpret $\MM$ as the topological invariant
associated with the first Chern character class $[F]$ in the second
cohomology (see below). With this we don't need to introduce magnetic
source terms into the homogeneous Maxwell equation but rather
interpret magnetic monopoles as a topological feature -- which one may
visualize as a Dirac string \cite{dirac31} or rather accept as a
feature of a $U(1)$-bundle (see figure \ref{dirac}). We choose the
nomenclature \emph{quasi-electric} and \emph{-magnetic} to remind us
of the analogies with electromagnetism. Since these definitions are
general and not restricted to theories of gravitation, we do not
choose the names \emph{gravi}-electric and -magnetic.

\begin{figure}\center
\input{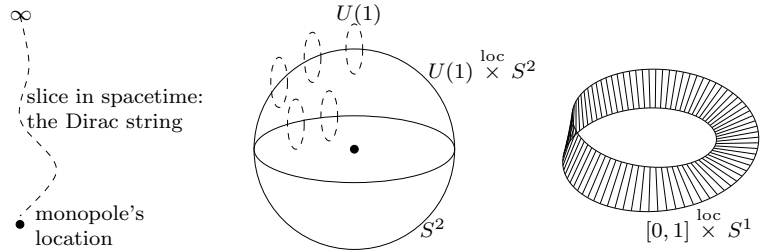}
\caption{
  The field strength of the Dirac monopole \cite{dirac31} $F=p\, d\O =
  p\, \sin\t\, d\t \w d\p$ has no global potential $A$ with $F=dA$.
  Dirac concluded that such a monopole must have a \emph{string}
  (slice in spacetime) attached to it. If we slice spacetime along the
  negative $z$-axis, say, $F$ has a regular potential $A=p\,
  (1-\cos\t)\, d\p$. Alternatively, electromagnetism may be formulated
  as a gauge theory on a $U(1)$ bundle over spacetime. Topologically
  the spacetime \emph{around} the (singular) monopole world path is
  $(\RRR^3_{\rm space} \setminus \{o\}) \times \RRR_{\rm time} \sim
  S^2$, where $o$ denotes the monopole's location. Hence, all field
  configurations may be classified topologically by investigating
  $U(1)$ bundles over $S^2$.  It turns out that an integer number (the
  magnetic charge) classifies all field configurations. The Moebius
  strip ($[0,1]$ bundle over $S^1$) allows to visualize a
  topologically non-trivial bundle.}
\Label{dirac}
\end{figure}

\subsection{Topological charges}\Label{Ctopcharge}

One principle of topology is comparing manifolds by continuously
deforming them. If two manifolds can continuously be deformed into
each other, they are said be homeomorph. In topology one is mainly
interested in the equivalence classes of homeomorph manifolds. It
turns out that there are three important ways of classifying
manifolds: First, by identifying all homeomorph manifolds with a set
of simplices that are glued together (homology). Second, by
considering those forms on the manifold that are closed but not exact
(cohomology). In some way (recalling the Stokes theorem) it is not
surprising that these two ways of classification are equivalent (de
Rham theorem). And third, by considering maps merging a topologically
well understood manifold (usually a $r$-sphere) into the manifold in
question (homotopy). We take \cite{naka} as a reference for topology.

For us the second way, i.e.\ considering the cohomology group $H^r(M)$
of $r$-forms over $M$ that are closed but not exact, is very
interesting. The Chern-Weil theorem enables to construct forms out of
the gauge field on a fibre bundle that are closed and of which the
exactness does \emph{not} depend on the gauge field. Such a form
represents \emph{one} element of the cohomology group independent of
the gauge field. Hence, this element of the cohomology group indicates
an \emph{invariant} (under gauge transformations) topological feature
of the bundle. One of these indicators, namely the first Chern
character class, may be used to define magnetic monopoles. We will
have a closer look on the Chern-Weil theorem below.

But also the third way of classifying the topology of $M$, i.e.\ 
considering the equivalence classes $\pi_r(M)$ of maps merging an
$r$-sphere into $M$, is very helpful. A theorem proved by Steenrod and
Pontrjagin (see, e.g., \cite{chanT} page 75) states that all
$G$-bundles over the base space $S^2$ can be classified by $\pi_1(G)$.
Since the world-path of a monopole is a singularity, the topology of
spacetime in the presence of a monopole is $\RRR^4 \setminus
\{$world-path$\} \sim S^2 \times \RRR^+ \times \RRR \sim S^2$, i.e.\ 
spacetime has the topology of $S^2$. Hence, all $G$-bundles over
spacetime can be classified by $\pi_1(G)$. In the case of
electromagnetism we have $G=U(1)$ and $\pi_1(U(1))=\ZZZ$, and all
gauge field configurations may be characterized by an integer number.
This tells us that, in general, there do exist topologically
non-trivial gauge configurations in electrodynamics.

Here, we define two topological charges:
\begin{align}
&\CC_I := \lim_{r\to\infty}\frac{1}{4\pi}\_{S^2(r)\times I} \oint \< A \w F \> 
   && \text{\em Chern-Simons charge,} \Label{cherncharge}\\
&\PP := \frac{g^2}{4\pi\hbar^2}\_{\RRR^4}\int \< F \w F \>
   && \text{\em Pontrjagin charge.} \Label{pontcharge}
\end{align}
Both charges are fruits of the Chern-Weil theorem which states that
these are topological invariants. We remind the reader of the
essential ideas of this theorem, for details see \cite{naka}.  First, consider the curvature $F \in
\L^2(P,\GG)$ on a principle bundle $P$ over the base manifold $M$ and
formulate polynomials $P(F)$ of this curvature. Then, search for such
polynomials that are invariant under the adjoint action of the
structure group $G$, i.e.\ $\forall g \in G:~ P({\rm Ad}_g F) = P(F)$.
Given such an invariant polynomial of $r$-th order, the Chern-Weil
theorem states the following:

(i) $P(F)$ is closed, i.e.\ $dP(F)=0$. Hence, we found an element of
the $2r$-th cohomology group $[P(F)] \in H^{2r}(M)$. Here, $[P(F)]$
denotes the equivalence class of all $2r$-forms that differ from
$P(F)$ only by an exact form. $[P(F)]$ is called \emph{characteristic
  class}. Note that each monomial in this polynomial is also
invariant.\\
(ii) If we have two curvatures $F$ and $F'$ on the same bundle it
follows that $[P(F)] = [P(F')]$. This means that the characteristic
class $[P(F)]$ is independent of $F$ and depends only on the topology
of the bundle. It is a topological invariant.\\
(iii) Since $P(F)$ is closed, we find a \emph{local} potential on a
subset $U$ of $M$: $P(F) = dQ\big|_U$. It follows that $[Q]$ is an
element of the $(2r\!\!-\!\!1)$-th cohomology $H^{2r-1}(\del U)$ and
is thus a topological invariant of $\del U$. $Q$ is called
\emph{Chern-Simons form}.

In fact, we find the invariant polynomials (or monomials) $P_1(F)=F$
and $P_2(F)=\<F \w F\>$, the first of which is called \emph{1st Chern
  character term} and the second \emph{1st Pontrjagin term}. We also
find the Chern-Simons form $\<A \w F\>$ of the 1st Pontrjagin term.

Hence, the 1st Chern character class $[F]$ is an element of the 2nd
cohomology. The integration of $F$ over a closed 2-plane $S^2$, i.e.\ 
the quasi-magnetic monopole charge $\MM$, thus leads to a number that
specifies the cohomology class.

Similarly, the Chern-Simons form $\< A \w F \>$ of the 1st Pontrjagin
term is an element of the 3rd cohomology and we need a closed 3-plane
for integration. In the case of a singular monopole world path in a
$U(1)$ bundle, a natural choice for this 3-plane is $S^2 \times U(1)$,
with $|U(1)|=\ell_5$. The integration $\CC_{U(1)}$ of the Chern-Simons
form over this plane thus leads to a finite number classifying the
cohomology class. Analogously we have a second choice $I=\RRR_{\rm
  time}$ to form a 3-plane $S^2 \times I$. However, this plane is not
compact and will not lead to a finite number. We solve this problem by
restricting $I$ to a finite time interval $I_T$ with $|I_T|=T$. Still,
the 3-plane $S^2 \times I_T$ is not closed and, strictly speaking,
$C_{I_T}$ may not be considered a topological invariant. Thus we have
to act with some caution.

The 1st Pontrjagin class $[\<F \w F\>]$ is determined by the
integration of $\<F \w F\>$ over a 4-plane -- which we always consider
to be spacetime. We will apply this definition in the context of a
translational gauge theory, i.e.\ a geometry with torsion $T$. Thus,
it is very instructive to note that the `translational Pontrjagin
term' $\<T \w T\>$ is equivalent to the Nieh-Yan term
\begin{align}
{\cal N}=T^\a \w T_\a - R_{\a\b} \w \v^\a \w \v^\b
\Label{NY}
\end{align}
in the case of vanishing curvature. The Nieh-Yan term may be produced
by splitting the 5-dimensional Pontrjagin term of a deSitter-like
$SO(5)$ gauge theory (via some inverse Inou-Wigner contraction) into
the 4-dimensional $SO(4)$ Pontrjagin term and the rest. This rest
refers to the translations and is, in fact, the Nieh-Yan term
(\ref{NY}). This was illuminated by Chandia and Zanelli
\cite{chandiaZ}.

\subsection{Elementary charges}\Label{Celementary}

One of the most beautiful things in physics is the success of particle
physics in classifying particles with the help of representation
theory for groups. This algebraic approach simply postulates that
objects in nature must be an element of a representation of some
symmetry. Objects (particles or states) that are inseparable are
called \emph{elementary}.  This notion turns out to coincide with the
mathematical notion of \emph{irreducibility}.  Both mean
\emph{inseparable} without loosing the symmetry (or a faithful
representation of it).

With elementary charge we denote those invariants that classify a
particle, i.e.\ the irreducible representation the particle is an
element of. Such a classification can be performed by finding all
Casimir operators in the group algebra. These are polynomials of the
group generators and commute with every group element. Hence, their
eigenvalues, when applied on some particle field, are invariant under
all symmetry transformations.

The Poincar\'e group, for example, has the Casimir operators
\begin{align}
C_1 &:= P_\a P^\a \;, \\
C_2 &:= W_\a W^\a \quad
\text{with} \quad W_\a := \frac{1}{2}\, \e_{\a\b\g\d}L^{\b\g}P^\d \;.
\end{align}
Here, the translation operator $P^\a$ represents the particle
momentum, $L^{\a\b}$ are the generators of Lorentz rotations, and the
so-called Pauli-Lubanski vector $W^\a$ represents the particle spin.
If nature incorporates the Poincar\'e symmetry, all particles can be
classified by eigenvalues of $C_1$ (mass square) and $C_2$ (spin
square). The classification with respect to their mass is guaranteed
by the Dirac equation (for the Dirac spinor representation) or the
Klein-Gordon equation (for the scalar representation). All these
equations require the dimension $\hbar/\ell$ for the mass parameter.
(We take $c=1$.)

In general, if the Casimir operator $C$ is a polynomial of $r$-th
order of the group generators and if $\II^r$ is an invariant
eigenvalue of $C$, i.e.  $(\hbar^r C - \II^r)\, \phi = 0$ for some
eigenvector $\phi$, then we call $\II$ an elementary charge.  If we
assume that $C$ is built from generators with dimension $[\l_a]$, the
dimension of $\II$ is $[\II]=\hbar\, [\l_a]$. This leads to the
remarkable relation between the dimension of an elementary charge and
that of a monopole charge (cf.\ table \ref{Tdim}):
\begin{align}
[\II] = [1/\k]\; [\EE^a] \period
\Label{elvermon}
\end{align}

\subsubsection*{Three further comments on mass and electromagnetic charges}

(1) Electric charge may as well be understood as an elementary charge of the
single $U(1)$-generator $P_5$, which is, of course, a Casimir
operator. To see this, decompose the $u(1)$-valued connection 1-form
into $A=A^5\,P_5$ (with $A^5$ having the conventional dimension of
Weber). The generator $P_5$ acts trivially on non-charged functions
$P_5 \cdot \psi \equiv 0$ but has any charged function as eigenstate
$P_5 \cdot \psi = e\, \psi$ with the elementary charge $e$. The
covariant derivative applied on the wave function of an electron, say,
reads $D \psi = d\psi + A\cdot \psi = d\psi + A^5\, P_5 \cdot \psi =
d\psi + e A^5 \psi$, as we are used to write it. This coupling of the
elementary charge to the gauge field induces the electric monopole
character of the electromagnetic field. Hence, the electric charge
density $\r$ may be understood as elementary charge density. A
magnetic monopole character, though, cannot be induced be an
elementary charge since there exists no second, magnetic-type Casimir
operator. Hence, some $\r_{\rm mag}$ on the rhs of the inhomogeneous
Maxwell equation may merely be understood as a density of
\emph{topological defects}, but not as elementary charge density.

(2) The dimension of the mass parameter $[m]=\hbar/\ell$ may be called
\emph{phase$/2\pi$ per length}. In fact, the most obvious argument for
this interpretation is the point particle action $\int\!  m\, ds$. In
this picture, if you identify a world path with a strap, then mass is
the twist of this strap per length. Also note that $\l_c=\hbar/m$ is
the Compton \emph{wave} length of the particle.

(3) Since in 5D Kaluza-Klein space the electric charge $q$ is just as
well an eigenvalue of the Casimir operator of the \emph{translation}
along the 5th dimension, electric charge is very similar to mass. Just
as mass measures the \emph{horizontal} (spacetime) momentum, the
electric charge measures a \emph{vertical} (fibre) momentum. In fact,
Bleecker \cite{bleecker} defined electric charge as the `vertical
velocity' of a point particle path on a $U(1)$-bundle.

\section[Translational monopole charges]{Translational monopole charges in gauge theories of gravity}\Label{Cgravcharges}

We can now apply the charge definitions to analyze standard solutions
of gauge theories of gravity for monopole charges. First, we
concentrate on a subclass of the Plebanski-Demianski class of
solutions including the Kerr-Newman and Taub-NUT solutions. For the
monopole analysis we formulate them as a solution of a translational
gauge theory of gravity, namely teleparallelism, and find
quasi-electric and quasi-magnetic monopoles in the gauge of some
translations, indeed. Later, we also investigate two solutions of the
Poincar\'e gauge theory.

Before we start we should point out that the following would hardly
have been possible without the use of the computer algebra system
Reduce and its supplementary package Excalc. The calculations for the
monopole analysis are rather straightforward but very extensive. At
the internet reference \cite{charge} we display the respective Reduce
files.

\subsection{The Kerr-Newman solution}

In polar coordinates $(t,r,\t,\p)$, the Kerr-Newman metric with mass
parameter $m$, Kerr parameter $j$, electric charge $q$, and magnetic
charge $p$ reads
\begin{align}
&g = \v^\0 \otimes \v^\0 - \v^\1 \otimes \v^\1
   - \v^\2 \otimes \v^\2 - \v^\3 \otimes \v^\3 \comma\\
&\v^\0 = \frac\QQ{\D}\, d\tau \comma
 \v^\1 = \frac{\D}{\QQ}\, dr \comma
 \v^\2 = \frac{\D}{\PP}\, \sin\t\, d\t \comma
 \v^\3 = \frac{\PP}{\D}\, d\s \comma \Label{coframe1} \\
& d\tau = dt - j \sin^2\t\, d\p \comma
  d\s   = (r^2 + j^2)\, d\p - j\, dt \comma \\
& \QQ^2 = r^2 - 2mr + j^2 + \frac{1}{4}\, (q^2+p^2)\comma
  \PP   = \sin\t \comma
  \D^2       = r^2 + j^2 \cos^2\t \period
\end{align}
This notation might confuse at first. It is the direct analogue of the
notation Plebanski and Demianski used in their paper \cite{plebD}.  It
has a clear structure and can easily be modified into other solutions
of the Plebanski-Demianski class. The metric solves the coupled
Einstein-Maxwell equations if we choose the electromagnetic potential
\begin{align}
A = \frac{1}{\D^2} \big( q\,r\, d\tau + p \cos\t\, d\s \big) \period
\end{align}
This potential is the analog of the potential $A = q/r\, dt + p
\cos\t\, d\p$ of an electric and magnetic charge in flat spacetime.
For the monopole analysis, we translate this solution into a 5D
Kaluza-Klein-type teleparallelism. This simply means that we add a 5th
dimension that represents the electromagnetic part of the theory:
\begin{align}
&g = \v^\0 \otimes \v^\0 - \v^\1 \otimes \v^\1
   - \v^\2 \otimes \v^\2 - \v^\3 \otimes \v^\3
   - \v^\5 \otimes \v^\5 \comma \Label{metric} \\
&\v^\5 = dx^5 
      + \frac{1}{\D^2} \big( q\,r\, d\tau + p \cos\t\, d\s \big) \period
\Label{coframe2}
\end{align}
The 5th covector $\v^\5$ represents the gauge of the 5th translation,
i.e.\ the electromagnetic gauge potential. The field strength of this
gauge theory is the torsion $T^\a = d \v^a$.  The configuration solves
the vacuum field equation $d H^a = 0$ of the teleparallelism theory.
Here, $H^a$ is the excitation of the translational gauge and is
composed out of the three irreducible pieces of $T^a$ such that the
theory is equivalent to 5D Einstein gravity:
\begin{align}
H^a=\frac{1}{\k} \h \Big({}^{(1)} T^a - 3 {}^{(2)} T^a + \frac{5}{2} {}^{(3)} T^a \Big) ~~~~ \text{or} ~~~~
H_\a = - \frac{1}{2} K^{\m\n} \w \eta_{\a\m\n} \;, \Label{choiceofh}
\end{align}
where $K^{\m\n}$ is the contortion. For details see \cite{hehlMMN}
or \cite{gron}.

The following charges for this gauge configuration are calculated by
the file {\tt kerrnut.exi} \cite{charge} with parameters
{\tt(m,j,q,p)}:
\begin{align}
\EE = -m\, \del_t - j\frac{\pi}{4}\, \del_\p + q\, \del_5 \comma
\MM = -p\, \del_5 \comma
\CC_{U(1)} = -p\, \ell_5 \comma
\CC_{I_T} = 0 \comma
\PP = 0 \;.
\Label{kerrcharges}
\end{align}
Consider $\EE$ and note that we have a quasi-electric monopole charge
$\EE^t=-m$ in the time translation, a quasi-electric monopole charge
$\EE^\p=-j\frac{\pi}{4}$ in the translation along $\del_\p$ (which is
actually a rotation and the charge represents an angular
momentum)\footnote{Usually, one associates a \emph{gravi-magnetic} or
  \emph{gravito-magnetic} effect with the gravitational field of the
  Kerr solution. This is sensible since the rotating mass produces a
  field that is in analogy to the magnetic field produced by rotating
  electrons. However, rotating mass is not an analogue to a magnetic
  \emph{monopole}. Instead, our calculation definitely proves that it
  is rather in analogy to an electric monopole -- but with respect to
  the gauge of translations along the Killing vector $\del_\p$.}, and
a (quasi-)electric monopole charge $\EE^5=q$ in the translation along
$\del_5$ (i.e.\ the $U(1)$ gauge of electrodynamics). In this solution
all Killing vectors carry quasi-electric monopole charges. In fact, it
seems quite plausible that the elementary charges of the three Casimir
operators (momentum square, Pauli-Lubanski square, and the 5th
translation) are the sources of the quasi-electric monopole charges of
the Killing vectors that correspond to these Casimirs in a stationary
geometry. As we are interested in dimensions, we find that the mass
parameter has dimension $[m]=\ell$, the angular momentum per mass unit
has dimension $[j]=1$, and, if we measure the length along the 5th
dimension in units of $\ell_5$, the electric charge has dimension
$[q]=\ell_5$. In the previous dimensional discussion of
electrodynamics, we defined $1/\ell_5=e/\hbar$ ~and~
$[1/\k]=e^2/\hbar=\hbar/\ell_5^2$. Hence, our results are consistent
with eq.\ (\ref{elvermon}): The dimension of the elementary charge
$[\II]=e=\hbar/\ell_5$ is equal to the coupling constant $[1/\k]$
times the dimension of the quasi-electric monopole charge
$[\EE^5]=[q]=\ell_5$. The same holds for the mass.

Considering $\MM$ we are not surprised that $\MM^5=-p$ is a
(quasi-)magnetic monopole charge of the 5th translation. The
non-trivial Chern-Simons form $\CC_{U(1)}$ confirms the topological
feature of magnetic monopoles in the $U(1)$-bundle.

\subsection{The Taub-NUT solution}

Let us turn to the Taub-NUT solution with mass parameter $m$, NUT
parameter $n$, and electric charge $q$. Within the previous notation,
i.e.\ with the coframe and metric defined in (\ref{metric},
\ref{coframe1}, \ref{coframe2}), the solution reads
\begin{align}
& d\tau = dt - 2n\cos\t\, d\p \comma d\s = (r^2 + n^2)\, d\p \comma\\
& p=0 \comma \QQ^2 = r^2 - 2mr -n^2 + q^2/4 \comma \PP
   = \sin\t \comma \D^2 = r^2 + n^2\period
\end{align}
The result of the monopole analysis has been calculated with the
program {\tt kerrnut.exi} \cite{charge} with parameters
{\tt(m,n,q)}:
\begin{align}
\EE = -m\, \del_t + q\, \del_5 \comma
\MM = -2n\, \del_t \comma
\CC_{U(1)}=0 \comma
\CC_{I_T} = -2n\, T \comma
\PP = 4n -\frac{q^2}{2n} \;.
\Label{nutcharges}
\end{align}
This clearly presents the NUT parameter $n$ as a quasi-magnetic
monopole charge of the time translation. Table \ref{Tcomp} gives
another illustration of these results.

\begin{table}[t]\center
\begin{tabular}{|c|c|}
\hline
\raisebox{-2ex}{}electric monopole & Schwarzschild solution\\ 
$A = -\frac{q}{r}\, dt$ & $Y^\0 = \v^\0 - dt = \left( \sqrt{1-\frac{2m}{r}} - 1 \right) dt \longrightarrow - \frac{m}{r}\, dt$\\
\raisebox{-2ex}{}$F=-\frac{q}{r^2}\, dt \w dr$ & $T^\0 \longrightarrow - \frac{m}{r^2}\ dt \w dr$\\
\hline
\raisebox{-2ex}{}magnetic monopole & Taub-NUT solution\\
$A = p (1-\cos\t)\, d\p$ & $Y^\0 = \v^\0 - dt \longrightarrow 2n(1-\cos\t)\, d\p$ \\
\raisebox{-2ex}{}$F=dA=p\, d\O$ & $T^\0 \longrightarrow 2n\, d\O$\\
\hline
\end{tabular}
\caption{
  The table compares the electric monopole with the Schwarzschild
  solution and the Dirac monopole with the Taub-NUT solution. The
  gravitational solutions are presented in a teleparallel formalism.
  Arrows $\longrightarrow$ mean the limit $r\to \infty$. The analogies
  between the electro-magnetic field strength $F$ and the field
  strength of time translation $T^\0$ confirm our interpretation of
  the mass parameter $m$ and the NUT parameter $n$.  The
  identification of $\v^\0 - dt$ with the gauge potential of time
  translation $Y^\0$ takes soldering into account.}
\Label{Tcomp}
\end{table}

\section{Relating to other formalisms}\Label{Clynden}

In this short section we will display the relation of our analysis to
more conventional ones. Lynden-Bell et al.\ \cite{lyndenN}, e.g.,
wrote a detailed review on monopoles in gravity and also discussed the
magnetic nature of NUT-space. Their considerations are based on the
following definitions of the gravo-electric and -magnetic fields. They
point out that a time-like Killing vector is necessary for this
definition and hence they consider the general \emph{stationary}
metric (cf.\ \cite{lyndenN} eq (3.1))
\begin{align}
g = f^2\,(dx^0 - A_i dx^i)^2 - \g_{ij}\,dx^i\,dx^j\;, \Label{lynmetric}
\end{align}
where $i=1,2,3$ and $A_i$ and $\g_{ij}$ are arbitrary. Motivated by
the expression of the force on a test particle with rest mass $m_0$
and velocity $\vec v$ in this geometry (cf.\ \cite{lyndenN} (3.2))
\begin{align}
\vec f =
\frac{m_0}{\sqrt{1-v^2/c^2}}
\left[\Big(-\frac{1}{f}\; \vec\nabla f\Big) + \frac{v}{c} \times \Big(f\, {\rm curl}\, \vec A \Big) \right]\Label{lynforce} \;,
\end{align}
and its formal analogy to the electromagnetic Lorentz force, they
define the gravo-electric and -magnetic fields as (cf. \cite{lyndenN}
(3.3,3.4))
\begin{align}
\vec E &:= -\frac{1}{f}\; \vec\nabla f \;, \Label{lynE}\\
\vec B &:= {\rm curl}\; \vec A\;. \Label{lynB}
\end{align}
We can now give another interpretation of these definitions by
reproducing them in our teleparallel formalism. The metric
(\ref{lynmetric}) is replaced by the coframe with
\begin{align}
\v^\0 = f\;(dx^0 \!-\! A)\;,
\end{align}
together with three spatial covectors $\v^i$ that are of no further
interest. We introduced the space-like 1-form $A=A_i\, dx^i$. Since in
table \ref{Tcomp} we notice a close relation between Newton's force
and the time component of torsion $T^\0$, we calculate
\begin{align}
T^\0 &= d \v^\0 = df \w (dx^0 \!-\! A) - f\; dA \feed
&= \frac{1}{f}\, df \w \v^\0 - f\; dA\;.
\end{align}
Following the conventional space-time decomposition of the
electromagnetic force we split this field strength of time translation
into an electric and magnetic part:
\begin{align}
T^\0&=-\big( E \w \v^\0 + B \big)\Label{ttt} \;,\\
E&:= - \frac{1}{f}\; df \;,\\
B&:= f\; dA\;.
\end{align}
Thereby we reproduced the definitions (\ref{lynE},\ref{lynB}) up to
the factor $f$ in $B$. However, looking at the force (\ref{lynforce})
it seems more consistent to include this factor $f$ in $B$ in order to
arrive at the conventional expression for the Lorentz force. We
conclude that the conventional formalism presented by Lynden-Bell et
al.\ is equivalent to our investigation in monopoles in the
\emph{time-component} of torsion $T^\0$. However, their formalism is
non-covariant at its very basis, it is insufficient to discuss
monopole charges in other translations (e.g.\ the Kerr parameter as
quasi-electric monopole charge in the translation along $\del_5$), and
it does not allow to identify quasi-magnetic charges with Chern-Simons
charges in the way we did. Finally, we cite the interesting statement
of Rindler \cite{rindler} section 8.12 according to which the minus
sign in (\ref{ttt}) -- which is the only difference to the
electromagnetic paradigm -- is due to the \emph{attractive} nature of
the gravitational force.

\section{Summary and discussion}\Label{Cchargedisc}

\begin{table}[t]\center
\begin{tabular}{|c||c|c|c|}
\hline
Casimir operators &Killing vectors &quasi-electric &quasi-magnetic \\
($\sim$ elementary charges) & &monopole charges &monopole charges\\
\hline
$C_1:=P_\m P^\m$ & $\del_t$ & $m$ & $n$ \\
$C_2:=W_\m W^\m$ & $\del_\p$ & $j$ & ($a$?) \\
$P_5$ & ($\del_5$) & $q$ & $p$ \\
\hline
\end{tabular}
\caption{
  The correspondence between Casimir operators, Killing vectors, and
  monopole charges in the Plebanski-Demianski class of solutions. The
  three columns to the right refer to a \emph{stationary} (and
  spherically symmetric) geometry. We have the Schwarzschild mass
  parameter $m$, Taub-NUT parameter $n$, Kerr parameter $j$,
  acceleration parameter $a$, electric charge $q$, and magnetic charge
  $p$.}
\Label{Tcharges}
\end{table}
The main results of this article are the dimensions summerized in
table \ref{Tdim}, the charge definitions
(\ref{elcharge}-\ref{pontcharge}), the dimensional relation
(\ref{elvermon}) between elementary and monopole charges, and the
explicit presentation of the charges (\ref{kerrcharges}) and
(\ref{nutcharges}) for the Kerr-Newman and Taub-NUT solution in the
teleparallel formulation, respectively. Table \ref{Tcharges}
summarizes the interpretation of these charges. All this has only been
possible because of the gauge theoretical formulation of gravity and
stresses the analogies between internal and external gauge theories.
Finally, we want to emphasize the following points:

(1) As we discussed in section \ref{Clynden}, the (gravo-) electric
and magnetic nature of the Schwarz\-schild and Taub-NUT solution,
respectively, can also be pointed out in the Riemannian formulation of
gravity. However, in the teleparallel formalism we arrived to recover
the Schwarzschild mass parameter and the NUT-parameter as monopole
charges \emph{of the time-translation}. First, this explains why
Lynden-Bell et al.\ need to assume a time-like Killing vector for
their definitions of gravo-electric and -magnetic fields, and second,
this uncovers the analogy between those charges and charges of other
translations, namely those along $\del_\p$ and $\del_5$. Furthermore,
our definitions (\ref{elcharge},\ref{magcharge}) have the advantage to
be covariant.

(2) We proved that in the Plebanski-Demianski class of solutions
\cite{plebD} (when reformulated as teleparallel solutions) the five
parameters $m$, $n$, $q$, $p$, and $j$ may be related to monopole
charges. Unfortunately, we could not confirm the same for the
acceleration parameter $a$. The reason might be the topologically
non-trivial coordinate transformation eq (4.4) in \cite{plebD}.
However, for consistency we may expect that $a$ relates to a
quasi-magnetic charge of the translation along $\del_\p$. Assuming
this, we agree with Plebanski and Demianski on their ordering of the
parameters: The six parameters should be ordered as three pairs ($m$,
$n$), ($j$, $a$), and ($q$, $p$) each pair of which belongs to the
time translation, the translation along $\del_\p$, and the
$U(1)$-translation, respectively. In each pair the first parameter
denotes the quasi-electric charge and the second parameter the
quasi-magnetic charge of these translations.

{\bf Acknowledgments}

The author is grateful to Prof. F.W. Hehl for his continuous
encouragement and many critical remarks on the subject.


\begin{thebibliography}{99}\thispagestyle{empty}

\bibitem{bleecker} D.\ Bleecker: \emph{Gauge theory and variational principles}. Addison-Wesley, London (1981). (Global Analysis, Pure and Applied, series A, no.\ 1)


\bibitem{chanT} H.\ Chan, S.T.\ Tsou: \emph{Some elementary gauge theory concepts}. World Scientific, Singapore (1993). (World Scientific Lecture Notes in Physics {\bf 47})






\bibitem{naka} M.\ Nakahara: \emph{Geometry, topology, and physics}. Adam Hilger, Bristol (1990).


\bibitem{oraif97} L.\ O'Raifeartaigh: \emph{The dawning of gauge theory}. Princeton Univ.\ Press (1997).

\bibitem{rindler} W.\ Rindler: \emph{Essential relativity}. Springer-Verlag, New York, 2nd edition (1997).










\bibitem{chandiaZ} O.\ Chandia, J.\ Zanelli: Topological invariants, instantons and chiral anomaly on spaces with torsion. Phys.\ Rev.\ {\bf D55} (1997) 7580-7585, Los Alamos e-Print Archive {\tt hep-th/9702025}.

\bibitem{dirac31} P.A.M.\ Dirac: Quantized singularities in the electromagnetic field. Proc.\ Roy.\ Soc.\ Lond.\ {\bf A133} (1931) 60-72.

\bibitem{dirac48} P.A.M.\ Dirac: The theory of magnetic poles. Phys.\ Rev.\ {\bf 74} (1948) 817-830.

\bibitem{gron} F.\ Gronwald: Metric-affine gauge theory of gravity: I.\ Fundamental structure and field equations. Int.\ Jour.\ Mod.\ Phys.\ {\bf D6} (1997) 263-303.



\bibitem{hehlMMN} F.W.\ Hehl, J.D.\ McCrea, E.W.\ Mielke, Y.Ne'eman: Metric-affine gauge theory of gravity: Field equations, Noether identities, world spinors, and breaking of dilation invariance. Phys.\ Rep.\ {\bf 258} (1995) 1-171.




\bibitem{lyndenN} D.\ Lynden-Bell, M.\ Nouri-Zonoz: Classical monopoles: Newton, NUT space, gravomagnetic lensing, and atomic spectra. Rev.\ of Mod.\ Phys.\ {\bf 70} (1998) 427-445.
 









\bibitem{plebD} J.F.\ Plebanski, M.Demianski: Rotating, charged, and uniformly accelerating mass in general relativity. Annals of Phys.\ {\bf 98} (1976) 98-127.




\bibitem{utiyama} R.\ Utiyama: Invariant theoretical interpretation of interaction. Phys.\ Rev.\ {\bf 101} (1956) 1597-1607.

\bibitem{yangM} C.N.\ Yang, R.L.\ Mills: Conservation of isotopic spin and isotopic gauge invariance. Phys.\ Rev.\ {\bf 96} (1954) 191-195.

\bibitem{charge} Internet reference of this paper: \newline{\tt http://www.thp.uni-koeln.de/\~{}mt/work/1999charge/index.html}

\end{thebibliography}
\end{document}